\documentclass[global,twocolumn]{svjour}

\usepackage{graphics}

\journalname{appliedphysicsB}

\begin{document}

\title{Phase Fluctuations in Bose-Einstein Condensates}

\author{
D. Hellweg\inst{1} 
\and S. Dettmer\inst{1}
\and P. Ryytty\inst{1} 
\and J. J. Arlt\inst{1} 
\and W. Ertmer\inst{1} 
\and K. Sengstock\inst{1} 
\and D. S. Petrov\inst{2,3} 
\and G. V. Shlyapnikov\inst{2,3,4} 
\and H. Kreutzmann\inst{5} 
\and L. Santos\inst{5} 
\and and M. Lewenstein\inst{5}
}

\institute{ 
Institut f\"ur Quantenoptik, Universit\"at Hannover, Welfengarten 1,
30167 Hannover, Germany
%%%%%
\and FOM Institute for Atomic and Molecular Physics, Kruislaan 407,
1098 SJ Amsterdam, The Netherlands
%%%%
\and Russian Research Center Kurchatov Institute, Kurchatov Square,
123182 Moscow, Russia
%%%%
\and Laboratoire Kastler Brossel, Ecole Normale Sup$\acute e$rieure,
24 rue Lhomond, 75231 Paris Cedex 05, France
%%%%
\and Institut f\"ur Theoretische Physik, Universit\"at Hannover,
Appelstra\ss e 2, 30167 Hannover, Germany 
}

%\date{Received: 10.07.01 / Revised version: date}
\maketitle

\begin{abstract}

% Abstract
We demonstrate the existence of phase fluctuations in elongated
Bose-Einstein Condensates (BECs) and study the dependence of those
fluctuations on the system parameters. A strong dependence on
temperature, atom number, and trapping geometry is observed. Phase
fluctuations directly affect the coherence properties of BECs. In
particular, we observe instances where the phase coherence length is
significantly smaller than the condensate size.  Our method of
detecting phase fluctuations is based on their transformation into
density modulations after ballistic expansion. An analytic theory
describing this transformation is developed.
   
\end{abstract}
\section{Introduction}
\label{intro}

%--------- Introduction ----------
Bose-Einstein condensates of weakly interacting gases, such as alkali
atom vapours, constitute very well controllable macroscopic quantum
systems. For extremely low temperatures ($T\rightarrow 0$) the
condensate is well described by an effective macroscopic single
particle wavefunction, occupied by millions of atoms
\cite{Dalfovo}. The corresponding macroscopic phase is related to many
fascinating properties of BEC such as its coherence, superfluidity,
and effects known from superconductivity as the Josephson effect. In
particular, the coherence properties are essential for very promising
applications of BEC such as matter wave interferometry
\cite{InterMIT,coherenceNIST,coherenceM,Bragg_InterNIST,Hall,Torii,Waveguide}
or atom lasers \cite{Atomlasers} which rely on BEC as a source of
coherent matter waves. The phase coherence of condensates was shown by
imaging the interference pattern of two independent condensates which
were brought to overlap \cite{InterMIT}. Also using an interference
technique it has been found that a trapped BEC has a uniform spatial
phase \cite{coherenceNIST} and therefore the coherence length is just
limited by the size of the condensate. This result has also been
obtained by measuring the momentum distribution in the radial
direction of a cigar shaped condensate by a spectroscopic technique
\cite{coherenceMIT}. These experiments have focused on the coherence
properties of almost pure condensates, prepared at temperatures well
below the critical temperature. The temperature dependence of the
coherence of a BEC was studied in a 'double slit' experiment where an
atom laser beam was extracted from a cigar shaped BEC at different
radial positions \cite{coherenceM}. A reduction of interference fringe
visibility was observed for increasing temperature with the
interference pattern being reproducible and thus indicating that the
relative phase of the condensate fraction was not fluctuating
randomly.

Even though the uniform spatial phase of a condensate has been
confirmed in several experiments, it is not an obvious or even general
property of BEC at finite temperature.  The 'fragmentation' into
independent condensates with randomly fluctuating relative phase was
often discussed in the context of the nucleation process of a
condensate \cite{Kagan}. There, the equilibrium state of the system
was assumed to be a 'pure' condensate without phase
fluctuations. However, before the condensate is completely formed
phase fluctuations between different regions of the condensate are
expected even though density fluctuations are suppressed. But also for
the equilibrium state of a quantum system it is expected that
low-dimensional (1D and 2D) quantum gases differ qualitatively from
the 3D case with respect to statistical and phase correlation
properties \cite{Shlyapnikov1D,Shlyapnikov2D,BEC1D2D}. Low-dimensional
quantum degenerate gases, where the trap energy-level spacing is
larger than the interaction energy between atoms, have been
experimentally realized in atomic hydrogen \cite{exp_quasi2D}, Sodium
\cite{Ketterle1D}, and Lithium \cite{Salomon} but phase correlation
properties have not yet been experimentally investigated. Recently, it
was shown theoretically \cite{Shlyapnikov3D} that for very elongated
condensates phase fluctuations can be pronounced already in the
equilibrium state of the usual 3D ensemble, where density fluctuations
are suppressed. This was experimentally and theoretically investigated
in \cite{Dettmer}.  The phase coherence length in this case can be
much smaller than the axial size of the sample.
 
In this paper we present detailed experimental and theoretical studies
of phase fluctuations in elongated BEC. The dependence on the trapping
geometry, temperature, and the number of condensed atoms is
investigated. We demonstrate that the spatial phase of the condensates
undergoes random fluctuations with an average value determined by the
system parameters. In particular, our measurements show that the
average phase fluctuations increase with tighter radial confinement.
Thus, they are especially pronounced for very elongated geometries as
are used to study the transition from 3D to 1D degenerate quantum
gases \cite{Ketterle1D,Dettmer}.

BEC has recently been achieved in elongated micro-circuit geometries
in which radial trapping frequencies of tens of KHz or even MHz are
possible \cite{ChipBEC}.  Since this is more than an order of
magnitude bigger than the radial frequencies at which we have observed
phase fluctuations our results should be especially relevant for these
systems.  Phase fluctuations should also be considered if tight
confinement waveguides are used for BECs \cite{Waveguide} or guided
atom lasers beams, even though our results directly apply only to the
equilibrium state of the system.
 
Note, that phase fluctuations lead to a broadening of the momentum
distribution, which is the Fourier transform of the single particle
correlation function, defining the coherence length. Therefore, only
in the absence of phase fluctuations, a BEC constitutes a state where
ultimate control over the motion and position of atoms, limited only
by Heisenberg's uncertainty relation, is achieved.
 
Fluctuations of the phase of a Bose condensate are due to thermal
excitations and always appear at finite temperature. Their
experimental characterization constitutes a test of many-body theories
at finite temperature. If the wavelength of these excitations is
smaller than all dimensions of the condensate they have a 3D character
and do not lead to pronounced phase fluctuations. In elongated
condensates however, low energy {\it axial} excitations have
wavelengths larger than the {\it radial} size of the sample and
therefore acquire a pronounced 1D behaviour leading to axial phase
fluctuations, although density fluctuations of the equilibrium state
are still suppressed. The coherence properties can be significantly
altered as compared with previous observations. In particular, the
coherence length, i.e. the distance at which the single particle
correlation function falls to e$^{-1}$, can be much smaller than the
axial size of the condensate. This is not in contradiction to previous
coherence measurements since they were performed in rather spherical
traps \cite{coherenceNIST}, in the radial direction of cigar shaped
condensates \cite{coherenceM,coherenceMIT}, and at low temperature
\cite{coherenceNIST,coherenceMIT}.

Our method to study phase fluctuations is based on ballistic
expansion.  We show that the original phase distribution is mapped
into the density distribution during time-of-flight. The density
modulations after ballistic expansion are a direct measure of the
original phase fluctuations of the trapped condensate.

The paper is organized as follows. It starts with a theoretical
discussion of phase fluctuations including a detailed explanation of
our analytic theory \cite{Dettmer} for the appearance of density
modulations in the ballistic expansion. Then we present our
experimental studies demonstrating the existence of phase fluctuations
in BECs and investigating their dependence on system
parameters. Finally, alternative methods to study phase fluctuations
are discussed, including simulations of the interference of phase
fluctuating condensates.

\section{Theoretical description}
\label{theory}

\subsection{Phase fluctuating condensates in elongated 3D traps}
\label{subsection_theory1}

In a standard equilibrium situation in 3D traps the fluctuations of
density and phase of the Bose-Einstein condensate are only important
in a narrow temperature range near the BEC transition temperature
$T_c$ (critical fluctuations).  Outside this region, the fluctuations
are suppressed and the condensate is phase coherent. This picture
precludes the interesting physics of phase-fluctuating condensates,
which is present in 2D and 1D systems (see
\cite{Shlyapnikov1D,Shlyapnikov2D} and refs. therein).

In the Thomas-Fermi regime (see e.g. \cite{Dalfovo}), when the
nonlinear mean field interaction energy dominates the kinetic energy,
the situation can change for elongated BECs \cite{Shlyapnikov3D}. While
density fluctuations in equilibrium will remain suppressed due to
their energetic cost, the situation can be different for the phase
fluctuations.  In particular, the axial phase fluctuations can
manifest themselves even at temperatures far below $T_c$. Then, as the
density fluctuations are suppressed, the equilibrium state of the
system becomes a {\it condensate with fluctuating phase}
(quasicondensate) similar to that in 1D trapped gases
\cite{Shlyapnikov1D}.  Decreasing $T$ gradually reduces the phase
fluctuations.

\subsection{Description of the fluctuating phase}
\label{subsection_theory2}

In this section we describe the phase fluctuations along the lines of
Ref. \cite{Shlyapnikov3D}. Let us consider a BEC at $T=0$ in the
Thomas-Fermi regime, where the mean-field (repulsive) interparticle
interaction greatly exceeds the radial ($\omega_{\rho}$) and axial
($\omega_x$) trap frequencies. The density profile of the
zero-temperature condensate has the well-known shape
$n_0(\rho,x)=n_{0m} (1-\rho^2/R^2-x^2/L^2)$, where $n_{0m}=\mu/g$ is
the maximum condensate density, with $\mu$ being the chemical
potential, $g=4\pi\hbar^2a/m$, $m$ the atom mass, and $a>0$ the
scattering length.  Under the condition $\omega_{\rho}\gg\omega_x$,
the radial size of the condensate, $R=(2\mu/m\omega_{\rho}^2)^{1/2}$,
is much smaller than the axial size $L=(2\mu/m\omega_x^2)^{1/2}$.

The phase fluctuations can be described by solving the Bogoliubov-de
Gennes equations \cite{Stringari} describing elementary excitations of
the condensate.  One can write the total field operator of atoms as
$\hat\psi({\bf r})=\sqrt{n_0({\bf r})}\exp(i\hat\phi({\bf r}))$, where
$\hat\phi({\bf r})$ is the operator of the phase, and the density
fluctuations have been already neglected following the arguments
discussed above.  The single-particle correlation function is then
expressed through the mean square fluctuations of the phase (see, e.g.
\cite{Popov}):
\begin{equation}  
\label{opdm} 
\!\langle\hat\psi^{\dagger}({\bf r})\hat\psi({\bf
r}')\rangle\!=\!\sqrt{n_0({\bf r})n_0({\bf r}')} 
\exp\{-\langle[\delta\hat\phi({\bf r},{\bf r}')]^2\rangle/2\},\!
\end{equation} 
with $\delta\hat\phi({\bf r},{\bf r}')=\hat\phi({\bf r})-\hat\phi({\bf
r}')$. The operator $\hat\phi({\bf r})$ is given by (see, e.g., \cite{Shev})
\begin{equation}
\label{operphi}    
\hat\phi({\bf r})=[4n_0({\bf r})]^{-1/2}\sum_{j}f_j^{+}({\bf r})\hat a_j 
+h.c.,
\end{equation}
where $ \hat{a}_j$ is the annihilation operator of the excitation with
quantum number(s) $j$ and energy $\epsilon_j$, $f_j^{+}= u_j + v_j$,
and the $u,v$ functions of the excitations are determined by the
Bogoliubov-de Gennes equations.

The ``low-energy'' axial excitations (with energies
$\epsilon_{j}<\hbar\omega_{\rho}$) have wavelengths larger than $R$
and exhibit a pronounced 1D behavior. Hence, one expects that these
excitations give the most important contribution to the long-wave
axial fluctuations of the phase.  The solution of the Bogoliubov-de
Gennes equations for such low-energy axial modes gives the spectrum
$\epsilon_j=\hbar\omega_x\sqrt{j(j+3)/4}$ \cite{Stringari}, where $j$
is a positive integer. The wavefunctions $f_j^+$ of these modes have
the form
\begin{equation} 
\label{fpm} 
f_j^{+}({\bf r})=\sqrt{\frac{(j+2)(2j+3)gn_0({\bf r})}{4\pi
(j+1)R^2L\epsilon_j}}P_j^{(1,1)}\left(\frac{x}{L}\right),  
\end{equation}
where $P_j^{(1,1)}$ are Jacobi polynomials. Note that the
contribution of the low-energy axial excitations to the phase operator
(\ref{operphi}) is independent of the radial coordinate $\rho$.

\subsection{Numerical simulations}
\label{subsection_theory3}

In order to simulate numerically the effect of phase fluctuations
during the ballistic expansion of the condensate we replace the
operators $\hat a_j$ and $\hat a^{\dag}_j$ in Eq. (\ref{operphi}) by
complex Gaussian random variables $\alpha_j$ and $\alpha^{*}_j$, with
the correlation $\langle \alpha_j\alpha^{*}_{j'} \rangle
=\delta_{jj'}N_j$, where $N_j$ is the occupation number for the
quasiparticle mode $j$ for a given (small) temperature $T$. Such a
random phase reproduces correctly the phase correlations, which for
not too large $|x-x'|/L\le 0.4 $, behave as
\begin{equation}
\label{center}
\langle[\delta\hat\phi(x,x')]^2\rangle_T=\delta_L^2|x-x'|/L,
\end{equation}
where the quantity $\delta_L^2$ is given by
\begin{equation}
\label{deltaL}
\delta_L^2(T)=32\mu T/15N_0(\hbar\omega_x)^2, 
\end{equation}
and $l_\phi=L/\delta_L^2$ can be interpreted as a phase coherence
length.  It is the distance at which the phase factor of the
single-particle correlation function (Eq. \ref{opdm}) falls to its
$1/e$ value.  The condition $l_\phi/L=1$ determines then a
characteristic temperature $T_\phi=15(\hbar\omega_x)^2N_0/32\mu$,
where $N_0$ is the number of condensed atoms. For $T_{\phi}<T_c$,
which is the case for most of our measurements, one expects the regime
of quasicondensation for the initial cloud in the temperature interval
$T_{\phi}<T<T_c$ \cite{Shlyapnikov3D}.

\begin{figure}
\resizebox{0.45\textwidth}{!}{%
  \includegraphics{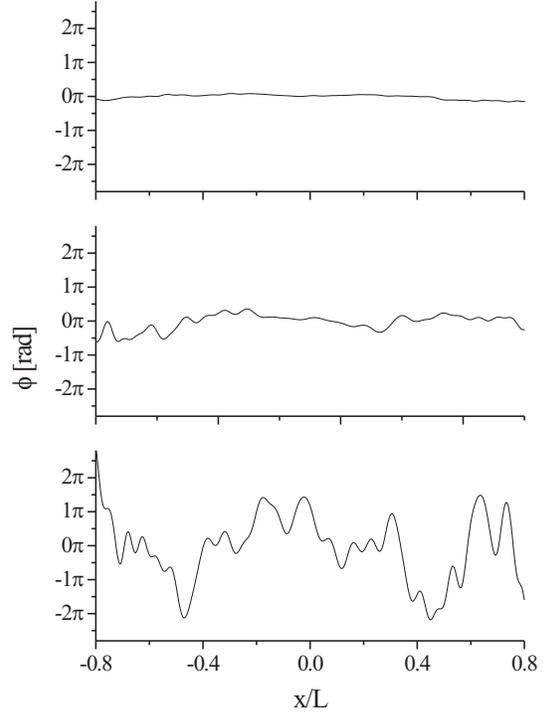}
}
%\vspace{5cm}       % Give the correct figure height in cm
\caption{Typical phase patterns for three different trap aspect ratios:
        $\lambda=10$ (top), $\lambda=100$ (middle), 
        and $\lambda=1000$ (bottom). For all cases $\omega_x=2\pi\times 14$ Hz, $T=0.6\; T_c$, 
        and $N=2\times10^5$. }
\label{figphase}       % Give a unique label
\end{figure}
 
Figure \ref{figphase} shows examples of simulated phase distributions
for various trap geometries parameterized by the ratio of trapping
frequencies $\lambda=\omega_\rho/\omega_x$.  Whereas for a trap with
$\lambda=10$ the fluctuations are rather suppressed, the phase of the
BEC fluctuates for a trap with $\lambda=1000$ by more than $2\pi$.
This dramatically changes the properties of the condensate, especially
in phase sensitive experiments.

The appearance of stripes in the process of ballistic expansion can be
understood qualitatively as follows. As mentioned above, within the
equilibrium state of a BEC in a magnetic trap the density distribution
remains largely unaffected even if the phase fluctuates
\cite{Shlyapnikov3D}.  The reason is that the mean-field interparticle
interaction prevents the transformation of local velocity fields
provided by the phase fluctuations into modulations of the
density. However, after switching off the external trap, the
mean-field interaction rapidly decreases and the axial velocity field
is then converted into a particular density distribution. We have
performed numerical simulations of the 3D Gross-Pitaevskii equation
(GPE) to understand quantitatively how phase fluctuations lead to the
build up of stripes in the density distribution.  We assumed that
initially (just before opening of the trap) the condensate had an
equilibrium density profile, which has been calculated using the
standard imaginary evolution of the GP equation. We have imprinted on
it a random fluctuating phase $\phi(x)$, as described above, and
evolved the condensate in free space.

\subsection{Analytic results}
\label{subsection_theory4}

Alternatively, the appearance of the stripes and their statistical
properties can be described analytically using the self-similar
solutions of the GP equation valid for the expanding cloud in the
Thomas-Fermi regime (see e.g. \cite{selfsimilar}).  In the following
we shall assume for simplicity that the condensate is an infinite
cylinder. This assumption is justified since the typical size of the
excitations is much smaller than the axial size of the
condensate. Therefore, at the end of the calculation, the unperturbed
density will be substituted by the corresponding local density. In
this way, the condensate without initial fluctuations evolve according
to the self--similar solution
\begin{equation}
\psi=\frac{\sqrt{n_0}}{b_\rho^2(t)}e^{i\phi_0},
\end{equation}
where $b_\rho^2(t)=1+\omega_\rho^2 t^2$,
$\phi_0=\frac{m}{2\hbar}\frac{\dot b_\rho}{b_\rho}\rho^2$, $\rho$ is
the radial coordinate, and $n_0$ is the Thomas--Fermi density profile.
The equations which determine the ballistic expansion in presence of
fluctuations can be obtained by linearizing around the self--similar
solution, for the density $n_0+\delta n$, and the phase
$\phi_0+\phi$. Introducing this expressions in the corresponding
Gross--Pitaevskii equation, we obtain:
%\begin{mathletters}
\begin{eqnarray}
\dot {\delta n}&=&\frac{1}{b_\rho^2}\hat{\cal
O}\phi-\frac{\hbar\nabla_x^2(n_0\phi)}{m}, \label{dn} \\
\-n_0\dot\phi&=&\frac{gn_0\delta n}{\hbar
b_\rho^2}+\frac{1}{4b_\rho^2}\hat{\cal O}\left(\frac{\delta n}{n_0}\right)-
\frac{\hbar\nabla_x^2\delta n}{4m}, \label{phi} 
\end{eqnarray}
%\end{mathletters}
where $\hat{\cal
O}=-(\hbar/m)(\vec\nabla_{\rho'}n_0\cdot\vec\nabla_{\rho'}+n_0\nabla_{\rho'}^2)$, and 
$\rho'=\rho/b_\rho$. By combining Eqs.\ (\ref{dn}) and (\ref{phi}), we obtain:
\begin{equation}
\ddot {\delta n}-\frac{gn_0\nabla_x^2\delta
n}{mb_\rho^2}+\frac{\hbar^2\nabla_x^4\delta n}{4m^2}=
\frac{\partial}{\partial t}\left (\frac{\hat{\cal O}\phi}{b_\rho^2}\right
)+\frac{\hbar}{4b_\rho^2m}
\nabla_x^2\hat{\cal O}\frac{\delta n}{n_0}.
\label{dn2}
\end{equation}
The last term at the rhs of Eq.\ (\ref{dn2}) can be neglected if
$\mu/\hbar\omega_\rho\gg 1$ (i.e. in the Thomas--Fermi limit). The
third term in the lhs of the equation (quantum pressure term) is for
short times smaller than the second one on the lhs, but becomes
comparable with it for times $t\sim (1/\omega_\rho)\sqrt{\mu
m/\hbar^2k^2}$. This becomes an important point as discussed below.
Averaging over the radial profile (employing the expansion in powers
of $\rho$ as discussed in Ref.\ \cite{Stringari}), we end up with the
equation:
\begin{equation}
\ddot
{\delta n}-\frac{\mu}{2mb_\rho^2}\nabla_x^2\delta
n+\frac{\hbar^2}{4m^2}\nabla_x^4\delta n=0.
\label{dn3}
\end{equation}
Expanding into the Bogoliubov eigenmodes of the system, then 
\begin{equation}
\hbar^2\ddot{\delta n_k}+\epsilon_k^2(t)\delta n_k=0
\end{equation}
where 
\begin{equation}
\epsilon_k(t)=\sqrt{\frac{\mu\hbar^2k^2}{2mb_\rho^2}+\frac{\hbar^4k^4}{4m^2}},
\end{equation}
is the Bogoliubov spectrum.

For short times $\omega_\rho t\
\raisebox{-0.8ex}{${\stackrel{\textstyle <}{\sim}}$\ } 1$ the phonon
part of the spectrum is dominant, i.e. $\epsilon_k(t)\simeq
\sqrt{\frac{\mu\hbar^2k^2}{2mb_\rho^2}}$.  The resulting equation can
be solved in terms of hypergeometric functions.  For larger times, the
free particle part of the spectrum dominates, i.e.
$\epsilon_k=\frac{\hbar^2 k^2}{2m}$, and the equation can be solved in
terms of Bessel functions. The two regimes may be matched, using the
asymptotic expansions of both hypergeometric and Bessel functions.  In
this way, we end up with the analytic expression for the relative
density fluctuations
\begin{equation}
\frac{\hat \delta n}{n_0}=2\sum_j \tau^{-(\epsilon_j/\hbar\omega_\rho)^2}
\sin \left ( \frac{\epsilon_j^2\tau}{\mu\hbar\omega_\rho} \right )\hat\phi_j,
\label{delta}
\end{equation}
where the sum extends over the axial modes 
$\epsilon_j=\hbar\omega_x\sqrt{j(j+3)/4}$, 
$\tau=\omega_\rho t$, and $\hat\phi_j$ is the contribution of the $j$--th 
mode to the phase operator in Eq.\ (\ref{operphi}).

\begin{figure}
   \resizebox{0.45\textwidth}{!}{\includegraphics{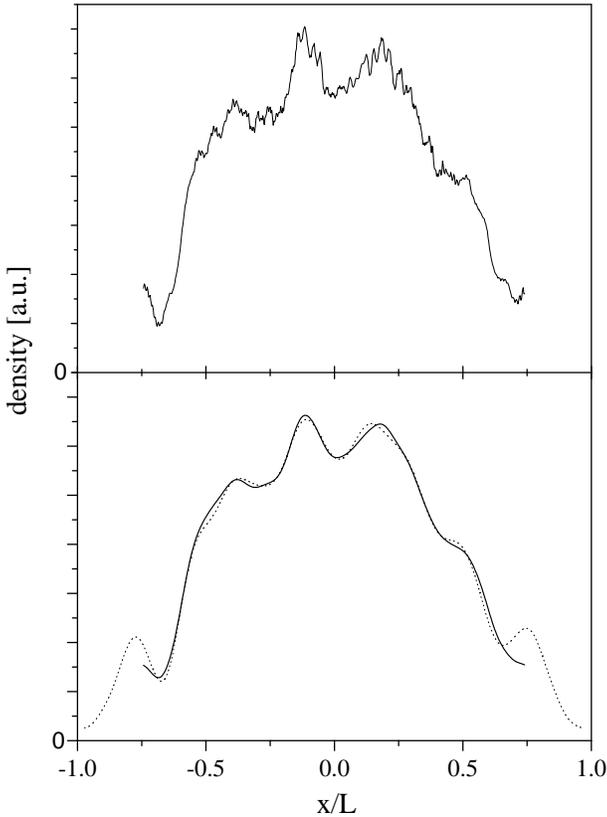}}
   \caption{Typical density profile after $25\, $ms time-of-flight.
        Top: numerical simulation without taking into account a
        limited experimental resolution.  Bottom: numerical simulation
        (dotted line) compared with the analytic theory (solid line),
        both taking into account an experimental resolution of
        $\sigma=3\,\mu\rm m$.  All profiles were calculated for
        $\omega_x=2\pi\times 14\, $Hz, $\omega_{\rho}=2\pi\times 508\,
        $Hz, $N_0=2\times 10^5$, and $T=0.5\, T_{c}$. For all the
        figures the same initial phase pattern was used.}
   \label{figden}       
\end{figure}

By substituting the operators by the corresponding c--numbers as
discussed above, we have checked that the analytical expressions agree
very well with the numerical results (see also Fig.~\ref{figden}).

\subsection{Averages}
\label{subsection_theory5}

From Eq.(\ref{delta}) one obtains a closed relation for the mean
square density fluctuations $\sigma^2$ by averaging $(\delta n/n_0)^2$
over different realizations of the initial phase. Taking into account
that $\langle (\alpha_j+\alpha_j^\ast)(\alpha_{j '}+\alpha_{j '}^\ast)
\rangle/4 = N_j\delta_{j j '}/2$, then the mean square fluctuations
are given by:
\begin{equation}
\label{C_Parameter}
\left \langle \left (
\frac{\delta n (x,t)}{n_0(x,t)} 
\right )^2 \right \rangle = 
\frac{T}{T_\phi} C(N_0,\omega_\rho, \omega_x, x,t)^2,
\end{equation}
where
\begin{eqnarray}
&& C(N_0,\omega_\rho, \omega_x, x,t)^2= \nonumber \\
&& \frac{1}{2} \sum_{j=1}^{\infty}\sin^2 \left(\frac{(j+3/2)^2}
{4\alpha(1-(x/L)^2)}\right) e^{-\left(\frac{\omega_x} 
{\omega_\rho}\right)^2 \frac{(j+3/2)^2}{2}\ln(2\omega_\rho
t)}
\nonumber \\
&&  
\left ( \frac{(j+2)(2j+3)}{j(j+1)(j+3)} \right ) (P_j^{(1,1)} 
\left(\frac{x}{L}\right))^2,
\label{Ccoef}
\end{eqnarray}
with $\alpha=\mu/\hbar\omega_x^2t$.

Using the quasiclassical average $(P_j^{(1,1)}(x))^2\simeq
4(1-x^2)^{-3/2}/\pi j$ and transforming the sum over $j$ into an
integral, for the central part of the cloud ($x\approx 0$) we obtain
\begin{eqnarray}
\left \langle \left (
\frac{\delta n (0,t)}{n_0(0,t)} 
\right )^2 \right \rangle = 
\left (
\frac{\sigma_{BEC}}{n_0}
\right )^2 =   \nonumber \\
\frac{T}{\epsilon
T_{\phi}}\sqrt{\frac{\ln\tau}{\pi}}
\left (
\sqrt{\!
1\!+\!
\sqrt{\!
1\!+\!
\left ( \frac{\hbar\omega_\rho\tau}{\mu\ln\tau} \right )^2
}
}
\!-\!\sqrt{2}
\right ).\!\!
\label{therm}
\end{eqnarray}
Note that Eq.~(\ref{therm})
provides a direct relation between the observed density fluctuations  and
temperature, and thus can be used for thermometry at very low $T$.

This closed expression would provide an accurate description of the
density fluctuations in the absence of any experimental
limitation. However, in practice, the observation of the density
fluctuations is limited by the spatial resolution of the experiment.
This fact can be easily taken into account in the calculations by
substituting $P_j^{(1,1)} (x/L)$ in Eq.\ (\ref{Ccoef}) by the
corresponding function convoluted by a resolution function
$\zeta(x)=e^{-x^2/\sigma^2}/\sqrt{\pi}\sigma$, where $\sigma$
characterizes the experimental resolution (Fig.~\ref{figden}).

Other experimental limitations can be easily taken into account by
following a similar procedure. In particular, a smoothing of the
observed density ripples is produced if the laser that integrates the
condensate column density in the absorption detection scheme is not
exactly parallel to the density stripes.  This effect can be easily
incorporated into the calculations, by filtering the density
distribution with a cut--off function in the Fourier spectrum of the
form
$\exp(ikRb_\rho(t)\theta)\sin(kRb_\rho(t)\theta)/kTb_\rho(t)\theta$,
where $\theta$ is the angle between the laser and the density
ripples. We return to this point in the experimental section.

\section{Experimental Results}
\label{experimental}

\subsection{Experimental Setup}
\label{subsection_setup}

%--------- Experimantal Setup ----------
The experiment was performed with Bose-Einstein condensates of
$^{87}$Rb atoms in the $|F\!=\!2, m_{F}\!=\!+2\rangle $ hyperfine
ground state.  As described previously \cite{bouncingBEC,Solitons} we
load a magneto-optical trap with a few times $10^9$ atoms from a chirp
slowed thermal beam. This is followed by a short period of subdoppler
cooling and optical pumping into the desired magnetic sublevel.  The
atomic cloud is then loaded into an Ioffe-Pritchard type (cloverleaf)
magnetic trap and finally adiabatically compressed to allow efficient
rf evaporative cooling. The fundamental frequencies of our magnetic
trap are $\omega_x=2\pi\times14\, $Hz and $\omega_\rho=2\pi\times365\,
$Hz along the axial and radial direction, respectively.  Due to the
highly anisotropic confining potential with an aspect ratio
$\lambda=\omega_\rho/\omega_x$ of 26 the condensates are already
elongated along the horizontal x-axis.
 
In order to study the dependence of phase fluctuations on the trapping
geometry we realized a wide range of radial confinement strengths.  To
allow for stronger radial confinement we used a holographically
generated blue detuned Laguerre-Gauss mode (TEM$_{01}^*$) laser beam
to form a combined magnetic and optical dipole potential trap
\cite{Waveguide}.  In this combined trap BECs were produced in a two
step evaporation procedure. The atoms were first cooled in the 'pure'
magnetic trap to a temperature slightly above the transition
temperature $T_c$. The optical dipole potential was then turned on
adiabatically, and finally the desired temperature was achieved by rf
evaporation in the combined potential.  On the other hand less
elongated condensates were produced by using a higher offset-field for
the magnetic trap leading to a weaker radial confinement.
 
Our measurements were performed for an axial trap frequency of
$\omega_x=2\pi\times14\, $Hz and radial frequencies $\omega_\rho$
between $2\pi\times138\, $Hz and $2\pi\times715\, $Hz corresponding to
aspect ratios $\lambda$ between 10 and 51. After rf evaporative
cooling to the desired temperature, we wait for 1 sec (with rf
'shielding') to allow the system to reach an equilibrium state. We
then switch off the trapping potential within $200\, \mu $s and wait
for a variable time-of-flight before detecting the atomic cloud by
resonant absorption imaging with the imaging axis perpendicular to the
long axis of the trap.

\subsection{Ballistic expansion measurements}
\label{subsection_balexpansion}

%--------- Figure1: TOF images ----------
%
\begin{figure}
\center{
   \resizebox{0.35\textwidth}{!}{\includegraphics{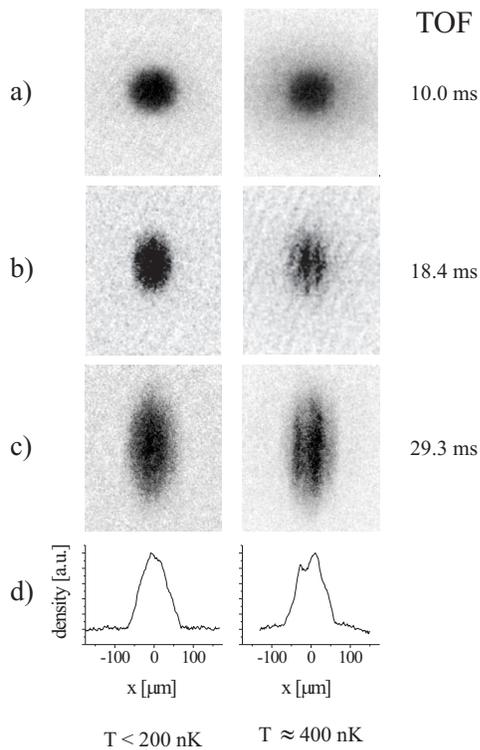}}
   \caption{(a)-(c): Absorption images for various times-of-flight and
      temperatures with $\omega_x=2\pi\times14\, $Hz and
      $\omega_\rho=2\pi\times365\, $Hz.  (d): Density profiles for the
      clouds displayed in (c) integrated along the radial
      direction. In the case of ${\rm T}<200\,{\rm nK}$ no thermal
      component was visible and the temperature was estimated to be
      $<$\,T$_c$/2.}
   \label{fig_exp1} }
\end{figure}

Figure \ref{fig_exp1} shows typical images of the ballistically
expanded clouds for various times-of-flight $t$ and temperatures
$T<T_c$. The usual anisotropic expansion of the condensate due to the
elongated trapping geometry is clearly visible.  The line density
profiles reflect the parabolic shape of the BEC density distribution.
As predicted by the theory, we also observe pronounced stripes in the
density distribution. On average these stripes are more pronounced
for high temperatures (right column of Fig.~\ref{fig_exp1}) and long
times-of-flight [Fig.~\ref{fig_exp1}(c)] indicating the build-up of
the stripes during the ballistic expansion.
%--------- TOF dependence ---------- 

\subsection{Experimental characterization of phase fluctuations}
\label{subsection_characterization}

%---------point diagramm ---------- 
In order to determine the dependence of phase fluctuations on the
experimental conditions we study the formation of the density
modulations as a function of temperature and trapping geometry.  All
measurements were performed for times-of-flight between 18 ms and 25
ms.  For a quantitative analysis we compare the observed density
distribution with the Thomas-Fermi distribution expected for a
condensate without fluctuations. For each image the observed density
distribution was integrated along the radial direction
[Fig.~\ref{fig_exp1}(d)] and then fitted by a bimodal function with
the integrated parabolic Thomas-Fermi density distribution and a
Gaussian for the thermal cloud. In Fig\ \ref{fig_exp1}(d) we show the
integrated density distribution close to the BEC region, thus the
thermal component is hardly visible. The standard deviations of the
experimental data from the fit were calculated in the central region
of the BEC (half width of full size), $\sigma_{\mbox{\tiny BEC}}$, and
in the thermal wings, $\sigma_{\mbox{\tiny T}}$.  The standard
deviation in the thermal wings characterizes our detection noise. To
account for particle number changes the standard deviations were
normalized to the fitted peak density $n_0$.

\begin{figure}
   \resizebox{0.45\textwidth}{!}{\includegraphics{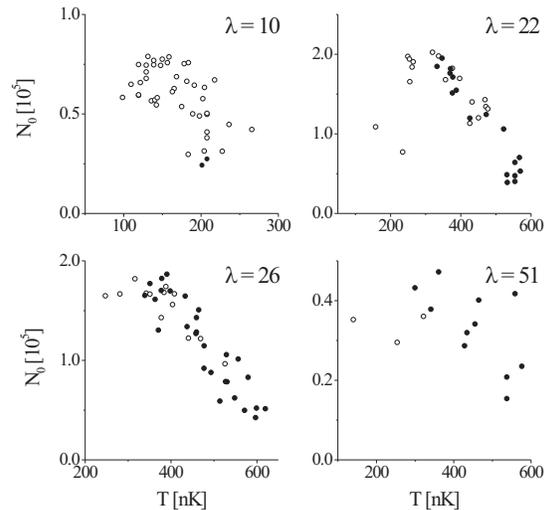}}
   \caption{Distribution of phase fluctuating condensates as a
       function of temperature and condensate number for four
       different trap geometries.  White points indicate the absence
       of detectable structures $(\sigma_{\mbox{\tiny BEC}}<1.5
       \sigma_{\mbox{\tiny T}})$, black points the existence of
       significant structures $(\sigma_{\mbox{\tiny BEC}}>2
       \sigma_{\mbox{\tiny T}})$.}
   \label{fig_exp2}
\end{figure}

Figure \ref{fig_exp2} shows the distribution of phase fluctuating
condensates as a function of temperature and condensate number for
various trap geometries. In this figure white points indicate the
absence of detectable structures in the density distribution
$(\sigma_{\mbox{\tiny BEC}}<1.5 \sigma_{\mbox{\tiny T}})$, whereas
black points indicate that significant structures larger than the
experimental noise level $(\sigma_{\mbox{\tiny BEC}}>2
\sigma_{\mbox{\tiny T}})$ were observed, i.e.\ the existence of phase
fluctuations could be clearly detected.  The temperature and particle
number of each condensate were determined by 2D fits to the absorption
images.  The temperature was determined from the width of a Gaussian
distribution fitted to the thermal wings, the corresponding condensate
number from the integral over the Thomas-Fermi part of a bimodal fit.
The statistical uncertainty in the temperature determination is
typically 15\,\% and less than 10\,\% for the particle numbers.

It is clearly visible in Fig.\ \ref{fig_exp2} that the number of
condensates showing detectable phase fluctuations is increasing with
the aspect ratio. Furthermore, the number of realizations without
detectable phase fluctuations is growing for low temperatures and high
particle numbers. Finally, there is no clear transition line between
the two regimes. There is rather a broad region in which we observe
both, condensates with and without detectable phase fluctuations
indicating the statistical character of the phase fluctuations.
Whereas for the weakest radial confinement with $\lambda = 10$, no
significant structures were observed outside the close vicinity of the
critical temperature, we detected pronounced phase fluctuations in a
broad temperature range in the case of our tightest trap $\lambda =
51$ [Fig.~\ref{fig_exp2}].  According to Eq.\ \ref{deltaL} phase
fluctuations can be reduced below any detection level for sufficiently
low temperatures. However, for high aspect ratio traps this requires
very low temperatures at high particle numbers in the condensate,
making it experimentally difficult to access.  For all traps the BECs
were produced by evaporating in the final potential except in the case
of the $\lambda = 10$ trap.  This trap was realized by evaporating in
the tighter $\lambda = 26$ trap and then adiabatically reducing the
radial confinement. Thus the measurements show that by changing the
trapping potential adiabatically we are able to decrease the amount of
phase fluctuations. The dynamics of the appearance and disappearance
of the phase fluctuations remains to be studied systematically in
future work.

%---------Fig 3: Sigma gegen T  ---------- 
\begin{figure}
   \resizebox{0.45\textwidth}{!}{\includegraphics{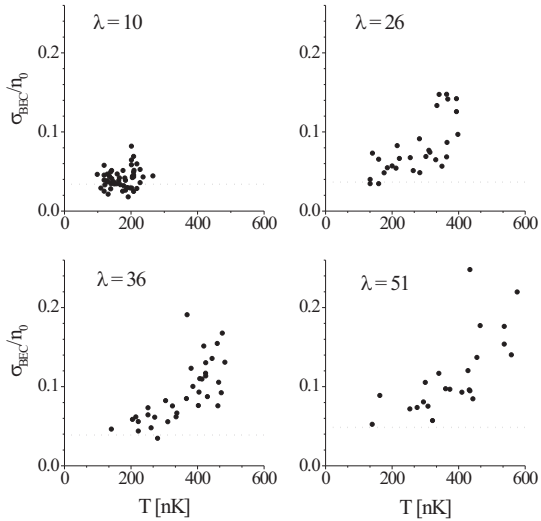}}
   \caption{Measurement of $\sigma_{\mbox{\tiny BEC}}/n_{0}$ versus
      temperature in four different trap geometries.  The dotted lines
      represent the average detection noise $\sigma_{T}/n_{0}$.  All
      data was taken for $T<T_{c}$.}
   \label{fig_exp3}
\end{figure}

Figure \ref{fig_exp3} shows the temperature dependence of
$\sigma_{\mbox{\tiny BEC}}$ for various trap geometries.  Since the
initial phase of a Bose condensate is mapped into its density
distribution after ballistic expansion, the quantity
$\sigma_{\mbox{\tiny BEC}}$ is a direct measure of the initial phase
fluctuations.  Note however, that this method reflects the
instantaneous phase of the BEC at the time of release and therefore
images taken at the same initial conditions can look significantly
different.  Indeed, we observe a large spread of our experimental data
(see Fig.~\ref{fig_exp3}). The scatter in these cases is mainly due to
the statistical character of the phase fluctuations as well as due to
the uncertainties in the temperature determination. For all traps the
highest temperature data correspond to values close to $T_c$
($T\approx0.9 T_c$) which is increasing for tighter confinement.  On
average the phase fluctuations continuously decrease with falling
temperature and get more pronounced with increasing aspect ratio.  The
reduction of the observed phase fluctuations for lower temperatures is
due to both, the reduced excitation spectrum at lower temperature and
the increasing number of condensed atoms.  Due to the loss of
particles caused by evaporation, lowering the temperature reduces the
total number of particles but the fraction of condensed atoms relative
to the total particle number is increased as shown in Fig.\
\ref{fig_exp2}.  It is not possible to determine a cut-off for the
phase fluctuations, rather they decrease until they cannot be resolved
below our noise limit. Hence all experiments with BECs at finite
temperature in tightly confining elongated potentials will be subject
to axial phase fluctuations.

\subsection{Evaluation of averaged phase fluctuations}
\label{subsection_evaluation}

%---------Fig 4: general behaviour  ---------- 
\begin{figure}
    \resizebox{0.5\textwidth}{!}{\includegraphics{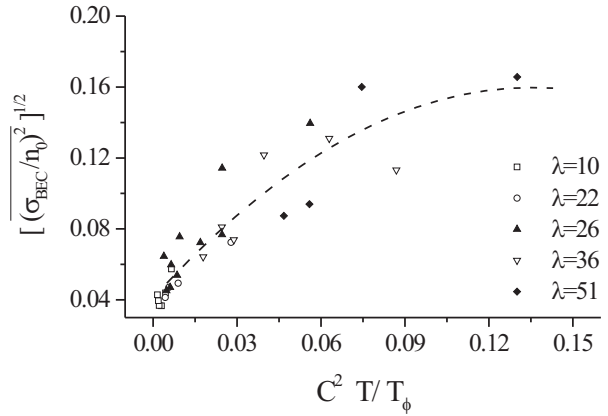}}
    \caption{Average standard deviation of the measured line densities
       $[\, \overline{(\sigma_{\mbox{\tiny BEC}}/n_{0} )^2}\,
       ]^{1/2}_{\mbox{\tiny exp}}$ as a function of temperature. The
       temperature-axis is scaled according to the expected dependence
       on the other experimental parameters (Eq. \ref{C_Parameter}).
       The dashed line is a fit of a square root function to the data
       reflecting the theoretical expected dependence.}
    \label{fig_exp4}
\end{figure}

Due to the statistical character of phase fluctuations every
experimental realization has a different phase distribution.
Therefore, we average the observed standard deviations for a small
range of experimental parameters to obtain general information about
the phase fluctuations and to compare our measurements to the
theoretical prediction.  According to Eq.\ (\ref{C_Parameter}) the
standard deviation of the fluctuations is expected to depend on the
square root of the temperature. This behaviour can not be observed in
Fig.\ \ref{fig_exp3} since the condensate number changes with
temperature. Therefore, we use that equation to scale out the
influence of condensate number, trapping potential, and
time-of-flight.  In figure \ref{fig_exp4} the measured standard
deviations are plotted as function of $T/T_\phi C^2$, where $T_\phi$
and C (C taken at x=0) are calculated from the measured number of
condensed atoms, trapping frequencies, and expansion times. The figure
shows that all data points fall on an universal square root shaped
curve. This means, that our experimental results follow the expected
dependence very well.

As a main result, with the direct link of the phase fluctuations in
the magnetic trap to the observed density modulations given by
Eq.~(\ref{therm}), our measurements confirm the predicted behaviour of
phase fluctuations in elongated BECs.

As explained above, a precise quantitative comparison needs to take
into account a reduced contrast due to the limited experimental
resolution and a possible tilt of the detection laser beam with
respect to the structures. If we account for a resolution of our
imaging system of $\sigma=3\mu\rm m$, the predicted standard
deviations are about a factor of 2 bigger than the observed
deviations. A possible reason might be a small tilt of the detection
laser beam. For our parameters a tilt of only $4^\circ$ reduces the
observed modulations by a factor of approximately 2.

Note, that our measurements were performed in elongated geometries but
our condensates still were in a 3D-regime in the sense that the
chemical potential was greater than the transverse level spacing.
Nevertheless, most of our measurements, which exhibit fluctuations
well above our noise level, correspond to the regime of
quasicondensation in which the phase coherence length is smaller than
the condensate size.  For instance, for $\lambda=51$, $T=0.5\, T_{c}$,
and $N_{0}=3\times 10^4$, one obtains $\mu=3.4 \hbar\omega_\rho$ and
$l_{\phi}\approx L/3$.

\section{Discussion of alternative methods}
\label{Discussion}

In principle, phase fluctuations may be observed by all experiments on
BEC which rely on its phase properties. Hence, various methods may be
used to study them.  We have presented a method which is based on the
transfer of phase fluctuations into density modulations after
ballistic expansion. A great advantage of our method is its
applicability to very different trap geometries and that the
sensitivity depends on the chosen time-of-flight t (see
Eq.~(\ref{therm})).  In principle for longer t the method becomes more
and more sensitive for initial phase fluctuations. On the other hand
the decreasing density reduces the signal-to-noise ratio leading to an
optimum time-of-flight for given parameters.  In our case a
time-of-flight of 25 ms allows us to measure phase fluctuations of
only $\delta_L^2 \approx\pi/7$.  Moreover, time-of-flight methods
constitute a standard experimental tool to study properties of
BECs. Therefore, understanding the formation of stripes in the density
distribution is of great importance.  It is instructive to discuss two
of the most successful methods used to study the coherence properties
of BEC, namely Bragg spectroscopy \cite{coherenceMIT} and
interferometry \cite{coherenceNIST}, with respect to the measurement
of phase fluctuations.
  
\subsection{Bragg spectroscopy}
\label{Discussion_a}

The velocity field of a condensate is proportional to the gradient of
its phase.  Therefore, phase fluctuations lead to a broadening of the
momentum distribution of a trapped condensate which can be measured
using Bragg spectroscopy. Since in elongated condensates phase
fluctuations are predominantly provided by axial excitations the
momentum distribution along the axial direction has to be measured.
The average momentum distribution $\cal P$ is given by the Fourier
transform (${\cal FT}$) of the integrated single-particle correlation
function $G({\bf r_a})\equiv \int d^3r
\!\langle\hat\psi^{\dagger}({\bf r})\hat\psi({\bf r+r_a})\rangle\!$
\cite{CCT_Cargese}. Using Eq.\ (\ref{opdm}) and the definition of the
phase coherence length we get
\begin{eqnarray}  
{\cal P}(p)&=&{\cal FT}\left[\exp^{-\frac{|{\bf
r_a}|}{l_\phi}}\int{d^3r \sqrt{n_0\left({\bf r}\right)n_0\left({\bf
r+r_a}\right)}}\right] \nonumber\\
&=&{\cal FT}\left[\exp^{-\frac{|{\bf r_a}|}{l_\phi}}\right]\star {\cal
FT}\left[\langle\sqrt{n_0\left({\bf r}\right)n_0\left({\bf
r+r_a}\right)} \rangle_r\right]
\end{eqnarray}
which is a convolution of the Fourier transform of the density
distribution and that of a Gaussian with the width of the phase
coherence length $l_\phi$. Thus, the first term is inversely
proportional to $l_\phi$, the second inversely proportional to the
condensate size.  In the absence of phase fluctuations the momentum
distribution is limited by the size of the condensate, i.~e.\ ,
Heisenberg-uncertainty limited.  In the regime of quasicondensates
where the phase coherence length is smaller than the condensate size,
the momentum distribution is dominated by the phase coherence length,
thus increasing the momentum width significantly with respect to the
Heisenberg limit.  However, for a trapped condensate the width of the
Bragg resonance is not only determined by this momentum width (Doppler
width) but also by the inhomogeneous density distribution, leading to
a spatially inhomogeneous mean field-shift of the resonance and
therefore broadening it.

We will now briefly estimate the relative contribution of the momentum
width and the mean-field width to the total linewidth of the Bragg
resonance.  For typical parameters of $N=10^5$, $w_\rho=2\pi \times
365\, $Hz, $w_x=2\pi \times 14\, $Hz the momentum width resulting from
the size of the condensate is $\Delta\nu_{\mbox{\small{size}}} =
\sqrt{21/8}\frac{2\hbar k}{2\pi mL} =57\, $Hz, where $k=2\pi /
\lambda$ is the absolute value of the wavevector of the Bragg
beams. For these parameters the mean field broadening is
$\Delta\nu_{\mbox{\small{mf}}} =\sqrt{8/147}\frac{\mu}{h}=536\, $Hz.
Hence the total width of the resonance is
$\Delta\nu=\sqrt{\Delta\nu_{\mbox{\small{momentum}}}^2 +
\Delta\nu_{\mbox{\small{mf}}}^2}=539\, $Hz.  Here, we have taken the
expressions for the widths from Ref.~\cite{coherenceMIT} and assumed
Bragg spectroscopy with counterpropagating beams since this leads to
the highest momentum resolution.  For a temperature $T=0.5 T_c$ the
phase coherence length is, according to Eq. (\ref{deltaL}),
$l_\phi\approx 1.4 L$ ($\delta_{\rm L}^2 \approx \pi/5$). Though this
increases the momentum width roughly by a factor of 2, it leads only
to a total linewidth of approximately 547 Hz. Thus, the contribution
of the momentum width to the total linewidth in this case is less than
2\%, making it very difficult to detect. In contrast, our measurements
show that the method of ballistic expansion is capable of detecting
even such small phase fluctuations.

\subsection{Interferometry}
\label{Discussion_b}

Interferometry has been successfully used to study and to characterize
the spatial phase of Bose condensates. Therefore, it is natural to
consider possibilities to extend such methods to determine the amount
of phase fluctuations in trapped atomic gases. In analogy to classical
optics, there are several methods to use interferometry to determine
the coherence properties of a condensate wave function. These include,
e.g., the double slit experiment with outcoupled atom laser
beams~\cite{coherenceM} and interference between two overlapping
condensates~\cite{InterMIT}. In the latter case, two spatially
separated condensates were prepared independently and their
interference was measured after a ballistic expansion. The results
indicated a flat phase distribution of the initial wavefunction for
the experimental parameters used in those measurements. In general,
interferometric measurements rely on the superposition of two
wavefunctions and the interpretation of the resulting fringe
pattern. The existence of a randomly fluctuating phase as shown in
Fig.\ref{figphase} complicates the interpretation.

To visualize possible experimental results, let us first consider the
following situation. A given phase fluctuating condensate is
coherently split in two equal parts giving one of them a small
additional velocity. Absorption images taken of the overlapping
condensates will show interference fringes. For condensates with a
constant phase the images would show regularly spaced lines. However
due to the phase fluctuations these lines will be shifted according to
the relative fluctuating phase, leading to a random, spatially varying
fringe spacing.  Similar to the ballistic expansion measurements these
images will be different from shot to shot and allow only for a
statistical evaluation.

\begin{figure}
   \resizebox{0.4\textwidth}{!}{\includegraphics{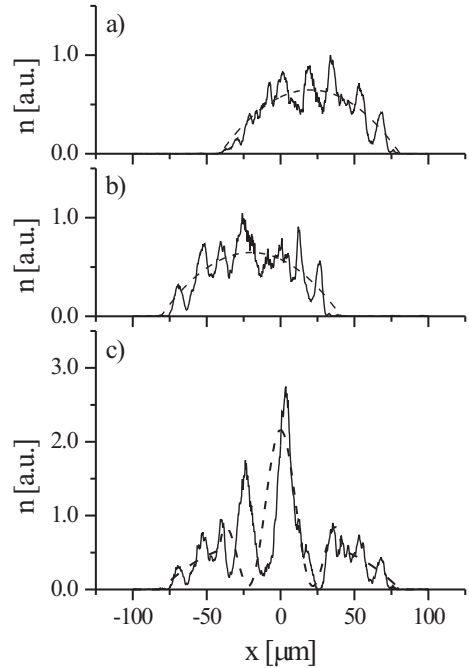}}
   \caption{(a, b) Examples of independently calculated density
        distributions of phase fluctuating condensates (solid line)
        for $\omega_\rho=2\pi\times715\, $Hz, $\omega_x=2\pi\times14\,
        $Hz, $N=5\times 10^4$ and $T=0.6 T_c$ compared to that of a
        condensate at $T=0$ (dashed line) for a time-of-flight of 15
        ms.  (c) Interference pattern of the superposition of the
        condensates shown in a) and b) with a relative displacement of
        $\Delta x=40\mu \rm m$.  }
   \label{fig_inter}      
\end{figure}

So far most interferometric measurements with BECs have been performed
after a ballistic expansion. An additional complication arises in
these experiments, since both density and phase of the sample evolve
during the free expansion time.  Let us assume that two independent
quasicondensates are brought to partial overlap after a time-of-flight
with zero relative velocity (using e.~g. Bragg pulses
\cite{Bragg_InterNIST,Torii,Waveguide}). At this point modulations
have appeared in the density distribution and the phase has evolved
from its initial random pattern. Both parts now contribute to the
formation of stripes in the interference pattern, making a direct
interpretation difficult.

Figure \ref{fig_inter} shows an example of an expected absorption
image for our experimental conditions. Both wavefunctions $\Psi_1$ and
$\Psi_2$ of the phase fluctuating condensates were calculated after a
time-of-flight of $15$ ms using the GP equation. The interference
pattern is given by $|\Psi_1(x)+\Psi_2(x+\Delta x)|^2$, where $\Delta
x$ is the displacement of the two quasicondensates.
Fig. \ref{fig_inter} a) and b) show the density distribution of the
two quasicondensates (solid line) displaced by $\Delta x=40\mu \rm
m\approx 2 l_\phi$ in comparison with the Thomas-Fermi distribution of
a condensate at $T=0$ (dashed line). Note, that the structures shown
there are the density modulations due to the phase fluctuations only.
The resulting interference pattern of these two independent
condensates is then displayed in Fig. \ref{fig_inter} c). At $T=0$ the
interference pattern is due to the small axial expansion velocity,
which leads to a quadratic phase. In the case of the quasicondensates
the profile looks considerably different, i.~e. the phase fluctuations
lead to a significant change of the experimental outcome. Due to the
difficult interpretation of the density distribution, interferometric
methods together with a long expansion time are not suited very well
for analysis of phase fluctuations. However, interferometric methods
that work with a short expansion time or even with trapped condensates
might be very useful for visualization of the spatial profile of phase
fluctuations. Such methods might include, e.g. preparation of two
independent quasicondensates with a small vertical separation. After a
short time-of-flight the radial expansion of the clouds leads to their
partial superposition, and consequently interference in the overlap
region. In the case of constant relative phase in the axial direction
horizontal stripes appear in the overlap region. Due to the phase
fluctuations the relative phase can, however, change as a function of
axial position, which is seen as bending of the interference fringes
in the vertical direction. Therefore, the relative phase profile of
the clouds can be imaged by measuring the shape of the interference
fringes.

\section{Conclusion}
\label{Conclusion}

We have presented detailed experimental and theoretical studies of
phase fluctuations in the equilibrium state of BECs. A strong
dependence on the trap geometry and temperature has been found in
agreement with the theoretical prediction. We have shown that phase
fluctuations are a general property of elongated condensates.  By
measuring the phase fluctuations and comparing the temperature with
$T_{\phi}$ we have demonstrated instances, where the phase coherence
length was smaller than the axial size of the condensate, i.e.\ the
initial cloud was a quasicondensate.  Our results set severe
limitations on applications of BECs in interferometric measurements,
and for guided atom laser beams.  Our experimental method combined
with the theoretical analysis provides a method of BEC thermometry.
Further studies of the effect of phase fluctuations, e.~g.\ on the
superfluid properties, will give additional insight to the behaviour
of degenerate quantum gases at finite temperature.\\

This work is supported by the {\it Deutsche Forschungsgemeinschaft}
within the SFB\,407 and the Schwerpunktprogramm "Wechselwirkungen
ultrakalter atomarer und molekularer Gase", and the European Science
Foundation (ESF) within the BEC2000+ programme.  DSP acknowledges
support from the Alexander von Humboldt Foundation, from the Dutch
Foundations NWO and FOM, and from the Russian Foundation for Basic
Research.

%\begin{figure}
%\resizebox{0.75\textwidth}{!}{%
%  \includegraphics{fig1.eps}
%\vspace{5cm}       % Give the correct figure height in cm
%\label{fig_exp1}       % Give a unique label
%\end{figure}
%\begin{figure*}
%\vspace*{5cm}       % Give the correct figure height in cm
%\end{figure*}
%

\end{document}